\documentclass[epsfig,12pt] {article}
\usepackage{amssymb}
\usepackage{amsmath}
\usepackage{amsfonts}
\usepackage{epsfig}
\usepackage{graphicx}

\include{definition}
\begin{document}

\author{ Navin Khaneja \thanks{To whom correspondence may be addressed. Email:navinkhaneja@gmail.com} \thanks{Systems and Control Engineering, IIT Bombay, Powai - 400076, India.}}

\vskip 4em

\title{\bf Electron conduction in solid state via time varying wavevectors}

\maketitle

\vskip 3cm

\begin{center} {\bf Abstract} \end{center}
In this paper, we study electron wavepacket dynamics in electric and magnetic fields. We rigorously derive the semiclassical equations of electron dynamics in electric and magnetic fields. We do it both for free electron and electron in a periodic potential. 
We do this by introducing time varying wavevectors $k(t)$.  In the presence of magnetic field, our wavepacket reproduces the classical cyclotron orbits once the origin of the Schr\"oedinger equation is correctly chosen to be center of cyclotron orbit. In the presence of both electric and magnetic fields, our equations for wavepacket dynamics differ from classical Lorentz force equations. We show that in a periodic potential, on application of electric field, the electron wave function adiabatically follows the wavefunction of a time varying Bloch wavevector $k(t)$, with its energies suitably shifted with time. We derive the effective mass equation and discuss conduction in conductors and insulators. 

\section{Introduction}
In this paper, we study electron wavepacket dynamics in the electric and magnetic fields by introducing time varying wavevectors $k(t)$. We do it both for free electron and electron in a periodic potential.  We derive the equation of group velocity of
electron wavepacket in electric and magnetic fields from first principle. 
We rigorously derive the semiclassical equations of electron in electric and magnetic fields.
In solid sate physics, the semiclassical equations of electron dynamics in electric and magnetic fields are well established and studied \cite{kittel}-\cite{cook}. The classical derivation goes something like the following \cite{kittel}. Suppose the electron wavepacket is assembled near a wavevector $k$, with group velocity $v_g = \frac{d \omega}{dk}$ with $\omega$ related to energy $\epsilon$ as $\epsilon = \hbar \omega$. The work done
on the electron by the electric field $E$ in time $\Delta t$ is 
\begin{eqnarray*}
\Delta \epsilon &=& -eE v_g \Delta t, \\
\Delta \epsilon &=&  \frac{d \epsilon}{dk} \Delta k = \hbar v_g \Delta k. 
\end{eqnarray*} Comparing above two equations we get $$ \hbar \frac{dk}{dt} = -e E, $$ which can be written for a more general force $F$ as
$\hbar \frac{dk}{dt} = F$, and with a electric-magnetic field it becomes $$ \hbar \frac{dk}{dt} = -e (E + v_g \times B). $$
Although above derivation is intuitive, it needs a rigorous justification from the first principles. In this paper, we provide a first principle derivation of electron wavepacket dynamics starting from Schr\"oedinger equation. We show how use of time varying wavevectors $k(t)$ in  Schr\"oedinger equation makes entire treatment transparent.

We then study wavepacket dynamics in periodic potential. We show that in a periodic potential, on application of electric field, the electron wave function and energies adiabatically follow the wavefunction and energy of time Waring $k(t)$. We derive the effective mass equation and discuss conduction in conductors and insulators.  We mention that the dynamics of electrons in homogeneous electric field has been previously studied in \cite{wannier}-\cite{krieger}. The paper is organized a follows. For expository reasons,  we first start in section \ref{sec:free} with free electrons and develop the use of the time varying wavevectors. In subsequent section \ref{sec:periodic}, we present the results of electron wavepacket dynamics in electric-magnetic fields in a periodic potential in a crystal. 

\section{Free Electrons}
\label{sec:free}

The free electron wavefunction is $\psi = \exp(i kx)$. The momentum is $\frac{\hbar}{i} \frac{\partial }{\partial x}$.
This gives the kinetic energy $\epsilon = \frac{p^2}{2m} =  \frac{\hbar^2 k^2}{2m}$, which for $\epsilon = \hbar \omega$ gives, 
\begin{equation}
 \omega(k) =  \frac{\hbar k^2}{2m}.  
\end{equation}

The dispersion is a parabola as shown below in  figure \ref{fig:dispersion} A.

\begin{figure}[htb!]
\centering
\includegraphics[scale = .35]{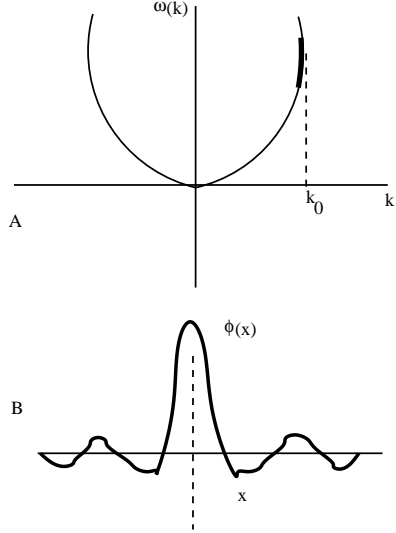}
\caption{Figure A shows the dispersion $\omega(k)$ vs $k$ for a free electron. Figure B shows a wavepacket centered at $k_0$. }
\label {fig:dispersion}
\end{figure}

Now, consider a wavepacket centered at $k_0$ shown the figure \ref{fig:dispersion} A, B. The packet takes the form

\begin{equation}
 \phi(x) = \frac{1}{\sqrt{N}} \sum_j \exp(i k_j x), \ \   \phi(x, t) =  \frac{1}{\sqrt{N}} \sum_j \exp(-i \omega(k_j) t) \exp(i k_j x),
\end{equation}

where $ \omega(k_j) = \omega(k_0) +  \omega'(k_0) \Delta k_j$ where $ \Delta k_j = k_j - k_0$. 
Denote $v_g = \omega'(k_0) =  \frac{\hbar k_0}{m}$, as the group velocity. Then 

\begin{equation}
\phi(x, t) = \frac{1}{\sqrt{N}} \exp(i (k_0 x - \omega(k_0) t)) \sum_j \exp(i \Delta k_j (x - v_g t)).
\end{equation}

The function $f(x) =  \frac{1}{\sqrt{N}} \sum_j \exp(i \Delta k_j x) =  \frac{2}{\sqrt{N}} \sum_j \cos(\Delta k_j x) $, is centered at origin with width $\propto (\Delta k)^{-1}$
as shown in figure \ref{fig:dispersion} B. Then 

\begin{equation}
| \phi(x, t) | = | f(x - v_gt)|,
\end{equation}
 the wavepacket moves with a group velocity $v_g$.

Now lets apply an electrical field $E$ in the $x$ direction at $t = 0$. Then the Schr\"oedinger equation is

\begin{equation}
i \hbar \frac{\partial \psi}{ \partial t} = \frac{1}{2m} (-i\hbar\frac{\partial }{\partial x})^2 + e E x \ \psi.
\end{equation}
The equation is satisfied by time varying wavevectors  $\exp(i k(t) x)$, where  $k(t) = k - \frac{eEt}{\hbar}$, with energy (dispersion) $\omega(k(t)) =   \frac{\hbar (k(t))^2}{2m} = \frac{\hbar (k-\frac{eEt}{\hbar})^2}{2m}$, so that the wavefunction becomes $$ \exp(-i \int_0^t \omega(k(\tau))\  d \tau) \exp(i k(t) x).$$ The initial wavepacket $\phi(x)$ evolves to $\phi(x, t)$, where,

\begin{equation}
 \phi(x) = \frac{1}{\sqrt{N}} \sum_j \exp(i k_j x), \ \   \phi(x, t) =  \frac{1}{\sqrt{N}} \sum_j \exp(-i  \int_0^t \omega(k_j(\tau)) d \tau ) \exp(i k_j(t) x).
\end{equation}

The group velocity 
\begin{equation}
  v_g(t) =   \frac{\hbar k(t)}{m} = \frac{\hbar (k-\frac{eEt}{\hbar})}{m} ; \ \ \frac{d v_g(t)}{dt} = -\frac{eE}{m}.
\end{equation}  

The electron wavepacket simply accelerates the way we know from classical mechanics. Being more pedagogical, we have 
\begin{eqnarray}
\nonumber \phi(x, t) &=&  \frac{1}{\sqrt{N}} \sum_j \exp(-i \int_0^t \omega(k_j(t))\ )\exp(i k_j(t) x) \\ &=& \frac{1}{\sqrt{N}} \exp(-i \int_0^t \omega(k_0(t))\ ) \exp(ik_0(t) x) \sum_j \exp(i \Delta k_j (x - \int_0^t v_g(\sigma) d \sigma  )).
\end{eqnarray}
The wavepacket evolves with instantaneous velocity $v_g(t)$.

The above method can be generalized to arbitrary potential. Consider the Schr\"oedinger equation

\begin{equation}
i \hbar \frac{\partial \psi}{ \partial t}  = (-\frac{\hbar^2}{2m} \frac{\partial }{\partial x^2} - e V(x) ) \psi. 
\end{equation}

\begin{figure}[htb!]
\centering
\includegraphics[scale = .35]{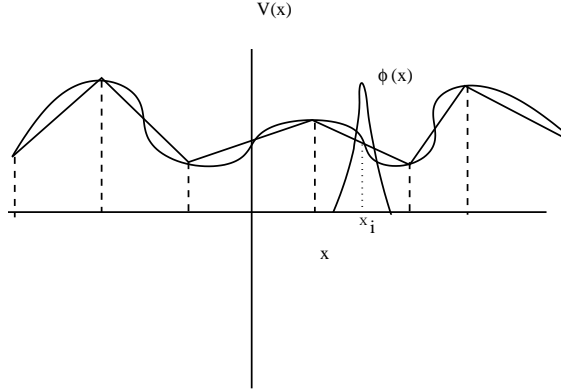}
\caption{Figure shows linear approximation of potential $V(x)$. The wavepacket $\phi(x)$ is confined to a cell.}
\label {fig:potential}
\end{figure}

We approximate the potential $V$ by piecewise linear potential such that $V(x) = V(x_i) + V'(x_i) \delta x$, where $\delta x = x - x_i$,
as shown in figure \ref{fig:potential}. We call these regions of linearized potential, cells. We can rewrite the potential in a cell as 
$V(x) = U(x_i) + V'(x_i) x$

We assume that the wavepacket has large $k_0$ such that $\Delta k \sim \sqrt{k_0}$ is large and therefore for the wavepacket, $\Delta x \sim (\Delta k)^{-1}$ is small so that it fits well within one cell. Then in this cell, the Schr\"oedinger equation takes the form

\begin{equation}
i \hbar \frac{\partial \psi}{ \partial t}  = (-\frac{\hbar^2}{2m} \frac{\partial }{\partial x^2} - e V'(x_i)x - e U(x_i) ) \psi. 
\end{equation} Since the wavepacket is confined to a cell, it evolution would be same if the potential we have was not only true in the cell but globally true. This is because the wavepacket doesn't know what the potential is outside the cell, its confined to the cell. Then lets solve the Schr\"oedinger equation with this potential assumed globally true and see how wavepacket evolves.

Then as before for the Schr\"oedinger equation is solved by wavevector $\psi = \exp(i k(t)x)$. Let $x(t)$ denote coordinates of center of wavepacket, then
\begin{equation} 
  k(t) = k +  \frac{e \int_0^t V'(x(\tau)) d \tau }{\hbar} , \ \ \omega(k(t)) =   \frac{\hbar (k + \frac{e \int_0^t V'(x(\tau)) d \tau}{\hbar})^2}{2m} - \frac{e \int_0^t U(x(\tau)) d \tau }{\hbar}.
\end{equation} The group velocity 

\begin{equation}
v_g(t) =  \frac{\hbar (k +  \frac{e \int_0^t V'(x(\tau)) d \tau}{\hbar})}{m} ; \ \ \frac{d v_g(t)}{dt} = \frac{eV'(x(t))}{m}.
\end{equation}

This is classical mechanics. Therefore at high energies where $k_0$ is large and wavepacket is well confined, i.e., over the packet width, the second order change of potential is small, $V''(x) \Delta x \ll V'(x)$. A linearized potential is a good approximation and evolution in quantum mechanics mimics classical mechanics.

We now consider both electric field $E$ and magnetic field $B \hat{z}$ in $z$ direction. To ease exposition we first consider $E=0$ case. The magnetic field is incorporated in the Schr\"oedinger equation by use of the gauge $A_x= -\frac{B}{2}y$ and $A_y = \frac{B}{2}x$. 

\begin{equation}
\label{eq:Aeqs}
i \hbar \frac{\partial \psi}{ \partial t} =  \frac{1}{2m} ((\frac{\hbar}{i} \frac{\partial }{\partial x} + eA_x)^2 + (\frac{\hbar}{i} \frac{\partial }{\partial y} + eA_y)^2)  \psi. 
\end{equation}

\begin{figure}[htb!]
\centering
\includegraphics[scale = .65]{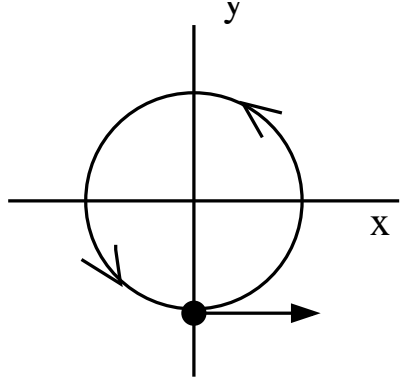}
\caption{The wavepacket is shown as bold dot which starts along $-y$ axis and circles origin.}
\label {fig:circle}
\end{figure}

We can write the Schr\"oedinger equation as

\begin{equation}
\label{eq:Aeqd}
i \hbar \frac{\partial \psi}{ \partial t} =  \frac{1}{2m} ((\frac{\hbar}{i} \frac{\partial }{\partial x})^2 + (\frac{\hbar}{i} \frac{\partial }{\partial y})^2 +  \frac{e\hbar B}{i} (x  \frac{\partial }{\partial y} -  y \frac{\partial }{\partial x}) + \underbrace{\frac{e^2B^2}{4} (x^2 + y^2)}_{V(x)})  \psi. 
\end{equation} The equation is solved by $\exp((ik_x(t)x + ik_y(t)y)$ with $\frac{dk}{dt}$ as follows. We denote ${\bf x}(t) = (x(t), y(t))$ and the cyclotron frequency $\omega_c = \frac{eB}{m}$. Then

\begin{equation}
\frac{dk}{dt}=  - \frac{\omega_c}{2} k \times \hat z - \frac{\omega_c^2 m}{4 \hbar} V'({\bf x}) = \frac{\omega_c}{2} k \times \hat z - \frac{\omega_c^2 m}{2 \hbar} {\bf x}(t). 
\end{equation}The dispersion $\omega(k(t)) = \frac{\hbar k(t)^2}{2m}$
and $v_g(t) =  \frac{\hbar k(t)}{m}$. Then

\begin{equation}
\label{eq:kcyclotron}
\frac{d v_g}{dt} =  - \frac{\omega_c}{2} v_g \times \hat z  - \frac{\omega_c^2}{2}  {\bf x}(t) = - \frac{\omega_c}{2}  v_g \times \hat z - \frac{\omega_c^2}{2}  ( {\bf x}(0) +  \int_0^t v_g dt ).
\end{equation}Suppose we start from ${\bf x}(0) = -( 0, \frac{\hbar k_x(0)}{m})$, with $k_y(0) =0$, then we have $v_g(t) =  \frac{\hbar k_x(0)}{m} (\cos \omega_c t, \ \sin \omega_c t)$ satisfies the above equation with ${\bf x} (t) =  \frac{\hbar k_x(0)}{m \omega_c} (\sin \omega_c t, \ -\cos \omega_c t)$ as shown in Fig. \ref{fig:circle}, which means

\begin{equation}
\label{eq:cyclotron}  
\dot{v}_g(t) = - \omega_c \ v_g \times \hat z = -\frac{e}{m} \  v_g \times B. 
\end{equation} If we differentiate Eq. (\ref{eq:kcyclotron} and \ref{eq:cyclotron}), we get same equation. In Eq. (\ref{eq:kcyclotron}), if ${\bf x}(0)$ is chosen so that $(v_g(0), \dot{v}_g(0))$ is same in Eq. (\ref{eq:kcyclotron} and \ref{eq:cyclotron}) then two equations will produce identical evolution of ${\bf x}(t)$. Therefore Eq. (\ref{eq:cyclotron}) stays valid provided the origin of Schr\"oedinger equation is at center of the cyclotron orbit.
We now consider both electric field $E$ and magnetic field $B \hat{z}$ in $z$ direction.
We can write  the Schr\"oedinger equation with ${\bf x}(t) = (x(t), y(t), z(t))$ as

\begin{equation}
\label{eq:AeqE}
i \hbar \frac{\partial \psi}{ \partial t} =  \frac{1}{2m} ((\frac{\hbar}{i} \frac{\partial }{\partial x})^2 + (\frac{\hbar}{i} \frac{\partial }{\partial y})^2 +  (\frac{\hbar}{i} \frac{\partial }{\partial z})^2 + \frac{e\hbar B}{i} (x  \frac{\partial }{\partial y} -  y \frac{\partial }{\partial x}) + \underbrace{\frac{e^2B^2}{4} (x^2 + y^2)}_{V(x)} + e E\cdot{\bf x})  \psi. 
\end{equation} The equation is solved by $\exp((ik(t) \cdot {\bf x})$ with, 

\begin{equation}
\frac{dk}{dt}=  - \frac{\omega_c}{2} k \times \hat z - \frac{\omega_c^2 m}{2 \hbar} {\bf x}(t) -\frac{e}{\hbar} E.
\end{equation}The dispersion $\omega(k(t)) = \frac{\hbar k(t)^2}{2m}$
and $v_g(t) =  \frac{\hbar k(t)}{m}$. Then

\begin{equation}
\label{eq:fundamental}
\frac{d v_g}{dt} =  - \frac{e}{2m} v_g \times B - \frac{\omega_c^2}{2}  {\bf x}(t) - \frac{e}{m} E.
\end{equation}  Recall the Lorentz force equation is

\begin{equation}
\label{eq:lorentz}
\frac{d v_g}{dt} =  - \frac{e}{m} v_g \times B - \frac{e}{m} E.
\end{equation} It is important to observe that Eq. (\ref{eq:fundamental}) and Eq. (\ref{eq:lorentz}) are different. Just differentiate both equations and we find they evolve different for same $(v_g(0), \dot{v_g}(0))$. They evolve same when $E =0$. Thus we claim  Eq. (\ref{eq:fundamental}) is more fundamental (In Eq. (\ref{eq:fundamental}), for $E=0$, we get cyclotron orbits if we choose our origin of Schr\"oedinger equation (\ref{eq:Aeqd}) correct). 

\section{Electrons in Periodic Potential}
\label{sec:periodic}

We now turn to electrons in a periodic potential created by periodic arrangement of ions in crystal. Once again to ease exposition, we first analyze in 
one dimension. Let $a$ be the lattice constant such that periodic potential $V(x) = V(x+a)$. We can Fourier decompose 
$$ V(x) = \sum_k V_k \exp(i kx). \ \ \ V_{-k} = V_k^{\ast}$$  The wavefunctions of the Bloch electrons \cite{bloch} take the general form 

$$ \psi_k(x) = \exp(i kx) u_{k, n}(x), $$ where $k$ is the crystal momentum lying in range $\frac{-\pi}{a} \leq k \leq \frac{\pi}{a}$,  $u_{k, n}(x+a) = u_{k, n}(x)$
is the periodic part of the wavefunction and $n$ is the band index. The corresponding energies  $\omega_{k, n}$ are eigenvalues of the system Hamiltonian

\begin{equation}
\label{eq:Hamiltrunc}
H = \left [ \begin{array}{ccccc} \ddots & \hdots & \hdots & \dots & \hdots \\ 0 & \frac{\hbar^2 (k + \frac{2 \pi}{a})^2}{2 m} +  V_0 & V_1 & \hdots & 0 \\
0 & V_{-1} & \frac{\hbar^2 k^2}{2 m} + V_0 & V_1 & 0 \\ 0 & 0 & V_{-1} & \frac{\hbar^2 (k -  \frac{2 \pi}{a})^2}{2 m} + V_0 & V_1 \\
0 & \hdots & \hdots  & \hdots & \ddots \end{array} \right ]. \end{equation} which for different $l$, couples free electron states 
$\exp(i (k + \frac{2 \pi l}{a}))$, with the periodic potential.

If $(\dots, a_{1}, a_0, a_{-1}, \dots)'$ is the eigenvector of $H$ in Eq. (\ref{eq:Hamiltrunc}) corresponding to energy 
$\epsilon_{k, n} = \hbar \omega_{k, n}$, then 

$$  u_{k, n}(x) = \sum_l a_l \exp(i \frac{2 \pi l}{a} x); \ \  \psi_{k, n}(x) = \sum_l a_l \exp(i (k + \frac{2 \pi l}{a}) x). $$

We now consider when only $n=0$ is populated, we suppress band index from now with $n=0$ implied. Now consider a wavepacket around $k=k_0$ written as 

$$ \phi(x) = \frac{1}{\sqrt{N}} \sum_j \psi_{k_j}(x) =  \frac{1}{\sqrt{N}} u(x)\sum_j \exp(i k_j x), $$
where $u_k(x)$ for $k$ around $k_0$ are approximated to be same $u(x)$. Then this wavepacket evolves as 

\begin{eqnarray*}
\phi(x, t) &=&  \frac{u(x)}{\sqrt{N}} \sum_j \exp(-i \omega(k_j) t ) \exp(i k_j x) \\ &=& \frac{u(x)}{\sqrt{N}} \exp(-i \omega(k_0) t) \exp(ik_0 x) \sum_j \exp(i \Delta k_j (x - v_g t )).
\end{eqnarray*} where $v_g = \frac{d \omega}{dk}|_{k_0}$. The wavepacket evolves with velocity $v_g$.

Now lets analyze the evolution of the wavepacket in the presence of an electric field $E$, say in $x$ direction. We again introduce time varying $k(t)$ with 

$$ k(t) = k + \frac{e A(t)}{\hbar}, \ \ \ A(t) = -Et. $$

then

\begin{equation}
\label{eq:timehamilton}
H(t) = \left [ \begin{array}{ccccc} \ddots & \hdots & \hdots & \dots & \hdots \\ 0 & \frac{\hbar^2 (k + \frac{2 \pi}{a} + \frac{e A(t)}{\hbar})^2}{2 m} + V_0 & V_1 & \hdots & 0 \\
0 & V_{-1} & \frac{\hbar^2 (k + \frac{eA(t)}{\hbar})^2}{2 m} + V_0 & V_1 & 0 \\ 0 & 0 & V_{-1} & \frac{\hbar^2 (k -  \frac{2 \pi}{a} + \frac{eA(t)}{\hbar})^2}{2 m} + V_0 & V_1 \\
0 & \hdots & \hdots  & \hdots & \ddots \end{array} \right ]. 
\end{equation}

Then observe 
\begin{equation}
\omega_{k, n} (t) = \omega_{k+\frac{eA(t)}{\hbar}, n} =  \omega_{k-\frac{eEt}{\hbar}, n}.
\end{equation}
This is how energies of $H(t)$ change when we apply electric field. To solve for time varying Schr\"oedinger 
equation we have to realize that for moderate $E$, the change of $H(t)$ is adiabatic and hence we really just follow the eigenvectors of $H(t)$. To see how this adiabatic following works, say at $t=0$, we are in the ground state say $X(0)$ (note $X(0) = (\dots, a_{1}, a_0, a_{-1}, \dots)'$, where $\psi_k(x) = \sum_l a_l \exp(i (k + \frac{2 \pi l}{a}) x).$) 
then $X(t)$ satisfies the Schr\"oedinger equation 

$$ \dot{X} = \frac{-i}{\hbar} H(t) X. $$ 

Lets diagnolize $H(t)$ as 

$$ H(t) = \Theta(t) \Lambda(t) \Theta'(t), $$ 

where $\Theta(t)$ is matrix of eigenvectors and $\Lambda(t)$ eigenvalues. 
Let us assume that $V_{-l} = V_l$ in Eq. (\ref{eq:timehamilton}), that is to say periodic potential is symmetric around origin. Then $H(t)$ is symmetric and $\Theta(t)$ a real rotation matrix. As we will see, this will ensure we don't have any geometric phases in our adiabatic evolution. Then we have $\dot{\Theta}(t) = \Omega(t) \Theta$, where $\Omega(t)$ is a skew symmetric matrix and we get  for $Y(t) =  \Theta'(t) X(t)$, 

\begin{equation}
\label{eq:adia} \dot{Y} =  (  \frac{-i}{\hbar} \left [ \begin{array}{cccc} \lambda_1 & 0 & 0 & 0 \\ 0 & \lambda_2 & 0 & 0 \\ \vdots & \hdots & \ddots & \vdots \\ 0 & 0 & 0 & \lambda_n \end{array} \right ] + \underbrace{\Theta' \Omega(t) \Theta}_{\bar \Omega} ) Y. 
\end{equation}where $\lambda_1$ is the smallest eigenvalue and so on. Note $Y(0) = (1, 0, \dots, 0)'$.

Now there is a gap between the lowest eigenvalue and higher eigenvalues which is atleast the band gap in range of eV. $\bar{\Omega}_{1j}(t)$ are comparatively small as we will show in following, and therefore they average out and $Y(t)$ simply evolves at $Y(t) = \exp(-\frac{i}{\hbar} \int \lambda_1(\sigma)) Y(0) =  \exp(-i \int \omega_k(\sigma)) Y(0) $ and we get 
that we just adiabatically follow the eigenvector of $H$ to get 

$$ \psi(t) =  \exp(-i \int \omega_k(\sigma) d \sigma) \psi_{k+\frac{eA(t)}{\hbar}} =   \exp(-i \int \omega_{k-\frac{eE\sigma}{\hbar}}d \sigma )  u_{k-\frac{eEt}{\hbar}}(x)\exp(ik(t) x). $$

To estimate how big are elements  $\bar{\Omega}_{1j}(t)$, observe 

\begin{eqnarray*}
\dot{H} &=&  \Theta(t) \dot{\Lambda}(t) \Theta'(t) + [ \Omega, H]  \\
\Theta'(t) \dot{H}  \Theta(t) &=&  \dot{\Lambda}(t) + [ \bar{\Omega}, \Lambda] \\
\| [ \bar{\Omega}, \Lambda] \| &\leq& \| \Theta'(t) \dot{H}  \Theta(t) \| = \|  \dot{H} \|.
\end{eqnarray*} Let $\bar{\Omega}_{\ast}(t) = \max \{ \bar{\Omega}_{1j}(t) \} $, then we have

\begin{equation}
\label{eq:bound}
(\lambda_1 - \lambda_2) \bar{\Omega}_{\ast}(t) \leq \| [ \bar{\Omega}, \Lambda] \| \leq \|  \dot{H} \|.
\end{equation}

Note we truncate $H$ in Eq. (\ref{eq:Hamiltrunc}) to finite size, because when $l$ on diagonal terms of $H$ become large the corresponding offdiagonals are rapidly truncated and in computing the ground state the submatrix with few small $l$ suffices. Usually $n=10$ is appropriate when $V_l$ are of order $eV$.

When we compute $\dot{H}$ we only have terms on diagonal and they are of the form $$ \frac{eE \hbar}{m} (k + \frac{eA(t)}{\hbar}), $$ then

\begin{equation}
\|  \dot{H} \| = \frac{eE}{\sqrt{m}} \sqrt{\sum_l \frac{\hbar^2 (k +  \frac{2 \pi l }{a}  + \frac{eA(t)}{\hbar})^2}{m}} = \frac{eE}{\sqrt{m}} \sqrt{2 tr(\bar{H})}.
\end{equation} where $\bar H$ is just $H$ without $V_0$ on the diagonal. Then  $ \bar{\Omega}_{\ast}(t)$ from Eq. (\ref{eq:bound}) satisfies ,

\begin{equation}
 \hbar \bar{\Omega}_{\ast}(t) \leq  \frac{\hbar eE}{\sqrt{m}(\lambda_1 - \lambda_2)} \sqrt{2 tr(\bar H)}.
\end{equation} Observe $(\lambda_1 - \lambda_2)$ is few $eV$ and $tr(\bar H)$ is also tens of  $eV$, but for field say $E = 10V/m$, the term

$$  \frac{\hbar eE}{\sqrt{m}} \sim \frac{10^{-33} \times 10^{-18}}{10^{-15}} = 10^{-36}. $$

$$  \frac{\hbar eE}{\sqrt{m}(\lambda_1 - \lambda_2)} \sqrt{tr(\bar H)} = 10^{-36} /\sqrt{10^{-18}} = 10^{-27} J = 10^{-8} eV. $$ 

Now we observe that $(\lambda_1 - \lambda_2)$ is of order $eV$ while  $\hbar \bar{\Omega}_{\ast}(t)$ is of order  $10^{-8} eV$, then 
in Eq. (\ref{eq:adia}), we have elements $ \bar{\Omega}_{1l}$ average out, giving us a pure adiabatic evolution.

It should be noted that in the equation

$$ \omega_{k} (t) = \omega_{k-\frac{eEt}{\hbar}}. $$

we treat $ k-\frac{eEt}{\hbar}$ in reduced Brillouin zone i.e,  for $b > 0$, $ k-\frac{eEt}{\hbar} = \frac{\pi}{a} + b $ is 
reduced to $ - \frac{\pi}{a} + b  \in [-\frac{\pi}{a}, \frac{\pi}{a}]$.

Now lets envisage a situation where we turn on a electric field in a conductor for some time $\tau$ and switch it off. Then  $A(t) = -\int_0^t E(\tau) d \tau $ rises from $0$ 
to a steady value $A$. This can be achieved as shown below in figure \ref{fig:solenoid}, where a conducting loop is pierced by a solenoid of area $a_r$. When current in the solenoid is turned on from $0$ to a steady state value, that creates a magnetic field $B(t)$ which goes from $0$ to a steady state value $B$ and hence establishes a transient electric field in the conducting loop and finally results in a steady $A$ in the loop such that 

$$ A = \frac{Ba_r}{2 \pi r}. $$

Then the electron wavefunction $\psi_k = \exp(ikx) u_k(x)$ is adiabatically transformed to $\exp(i(k+\frac{eA}{\hbar})x) u_{k + \frac{eA}{\hbar}}(x)$. The initial wavepacket 

$ \phi(x) =  \frac{1}{\sqrt{N}} u_{k_0}(x) \sum_j \exp(k_jx) $ is adiabatically transformed to $$ \phi(x) =  \frac{1}{\sqrt{N}} u_{k_0 + \frac{eA}{\hbar}}(x)  \sum_j \exp((k_j+\frac{eA}{\hbar})x). $$ If initially the wavepacket was moving with a group velocity $v_g = \frac{d \omega(k)}{dk}|_{k_0}$, now it moves with a group velocity
$v_g = \frac{d \omega(k)}{dk}|_{k_0 + \frac{eA}{\hbar}}$. This is shown in Fig. \ref{fig:wavepacket}, where $E(t)$ is transiently turned on and off and $A(t)$ reaches a steady state value $A$. The figure shows how $A$ in the conducting loop shifts the energy of a wavepacket and thereby changes the group velocity and hence accelerates the wavepacket. Lets estimate the shift $\frac{eA}{\hbar}$ for the example we have. Suppose solenoid has $N = 10$ turns per c.m. and carries a final current of $I = 1$ milli Ampere. Then it establishes a $B$ field of $B = \mu_0 N I$. Let $r$ be $.1$ m and $a_r = 1$ cm$^2$. Then we have $\frac{eA}{\hbar} = 10^4$m$^{-1}$. At half filling $k_0 = \frac{\pi}{2a}$, with $a = 1 A^{\circ}$, $k_0 \sim 10^{10}$m$^{-1}$ and is displaced $.0001\%$ by electric field.

\begin{figure}[htb!]
\centering
\includegraphics[scale = .35]{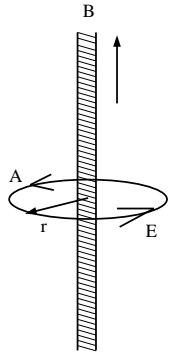}
\caption{Fig. shows a solenoid inside a loop conductor. As current in solenoid is turned on a magnetic field B is established inside the solenoid, which creates a $E$ in the loop that rises and decays and $A$ that goes from $0$ to a steady state value.}
\label {fig:solenoid}
\end{figure}

\begin{figure}[htb!]
\centering
\includegraphics[scale = .5]{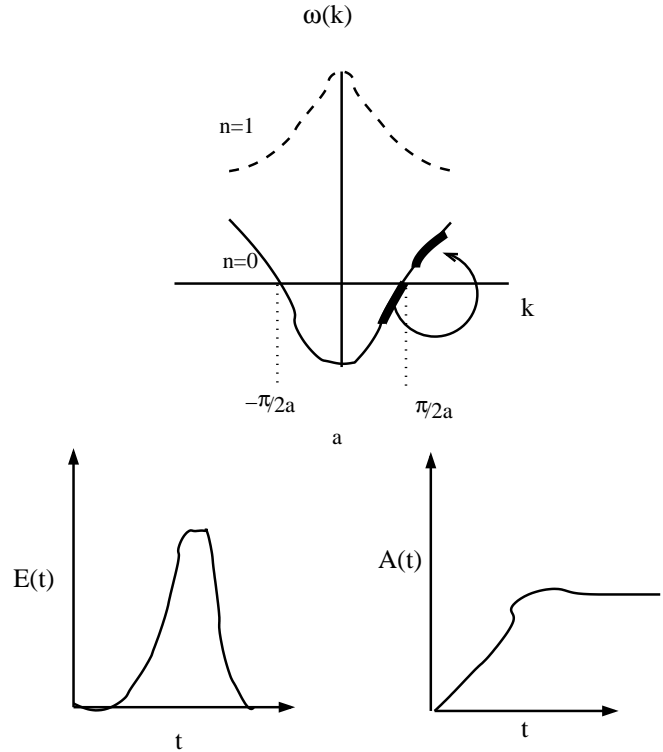}
\caption{Bottom figure shows how application of current in solenoid in Fig. \ref{fig:solenoid} establishes a transient $E$ and $A$ in the conducting loop which shifts the energy of a wavepacket and thereby changes the group velocity and hence accelerates the wavepacket as in top figure.}
\label {fig:wavepacket}
\end{figure}

\begin{figure}[htb!]
\centering
\includegraphics[scale = .5]{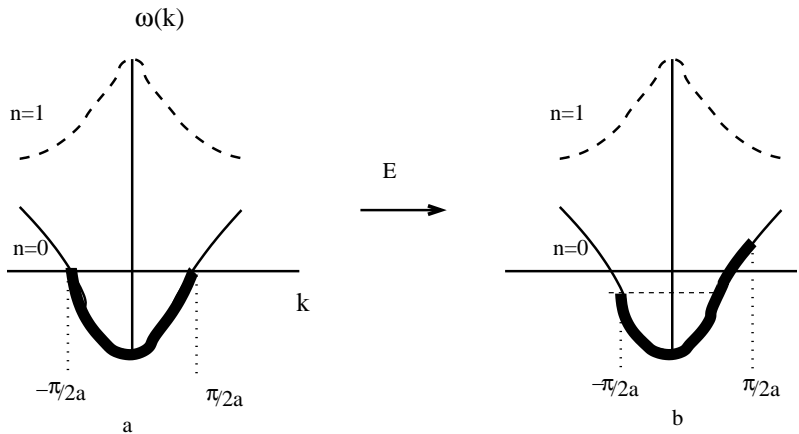}
\caption{Fig. a shows a half filled band. Fig. b shows how by application of electric field, the energies are shifted.}
\label {fig:band}
\end{figure}

\vspace{.2 in} 
We can now understand conduction in solids. Consider a half filled band in a conductor (say the conducting loop in above example) as 
shown in the figure \ref{fig:band}a. There is no net velocity of electrons. There are as many wavepackets moving in the positive direction as negative direction. While the wavepacket centered at $k_0$ has group velocity $v_g = \frac{d \omega}{d k} |_{k_0}$, there is a wavepacket at $-k_0$ 
with group velocity $-v_g$. There is no net current, though electrons are itenerant. Now say we apply a transient electric field as in solenoid example above which shifts the energy $\omega_k \rightarrow \omega_{k + \frac{e A}{\hbar}}$. Then we reach a configuration as shown in Fig. \ref{fig:band}b. Then we see we have excess of wavepackets with positive $v_g$. In the solenoid example we calculated the excess to be around $.0001 \%$. This gives net conduction and current. Therefore half filled bands conduct. Now imagine a insulator where band is completely filled, then all $k \in [-\frac{\pi}{a}, \frac{\pi}{a}]$ are occupied. After application of $E$ we have  $\omega_k \rightarrow \omega_{k + \frac{e A}{\hbar}}$, but because we are in the reduced Brillouin zone, nothing happens, as all energies are simply shifted, the energy of one wavepacket takes the value of another and so on. Although individual wavepackets are accelerated and deaccelerated the sum total of velocities is still zero. It is a common understanding that in a insulator all $k's$ are filled so electric field cannot accelerate a wavepacket and everything is stuck. In our picture, wavepackets are accelerated/deaccelerated but sum total of $v_g$ of all the wavepackets remains zero.

In an insulator as the electron wavefunction $\psi_k = \exp(ikx) u_k(x)$ is adiabatically transformed to $\exp(i(k+\frac{eA}{\hbar}) x) u_{k + \frac{eA}{\hbar}}(x)$,
when $k + \frac{eA}{\hbar} > \frac{\pi}{a}$, then putting it back in reduced Brillouin zone let $k + \frac{eA}{\hbar} = \frac{2 \pi}{a} - k'$, 
with $k' \in [\frac{-\pi}{a}, \frac{\pi}{a}]$, then observe $u_{k + \frac{eA}{\hbar}}(x)$ really means $\exp(-i \frac{2 \pi}{a} x) u_{-k'}(x)$.

Now in presence of an electric field we have,

\begin{equation}
\omega_{k} (t) = \omega_{k-\frac{eEt}{\hbar}}.
\end{equation}

Then the group velocity satisfies,

\begin{eqnarray}
v_g(t) &=& \frac{d \omega_{k-\frac{eEt}{\hbar}}}{d k}, \\ 
\frac{d v_g(t)}{dt} &=& \frac{d}{dk} \frac{d  \omega_{k-\frac{eEt}{\hbar}}}{dt} = \frac{d^2 \omega_{k-\frac{eEt}{\hbar}}}{dk^2} \frac{e E}{\hbar}
=  \hbar^{-2}\frac{d^2 \epsilon_{k-\frac{eEt}{\hbar}}}{dk^2} e E = -\frac{1}{m^{\ast}} eE, 
\end{eqnarray} where $m^{\ast} =  \hbar^{2}(\frac{d^2 \epsilon_{k+\frac{eEt}{\hbar}}}{dk^2})^{-1}$ is the effective mass.

For pedagogical reasons we have restricted to 1D , we can easily generalize the above to 3D. Let $\nabla'$ be gradient written as a column vector
and  $\nabla$  be gradient written as a row vector. Writing $v_g$ as column 3 vector,
\begin{eqnarray}
v_g(t) &=& \nabla_k' \omega_{k+\frac{eEt}{\hbar}},\\ 
\frac{d v_g(t)}{dt} &=& \nabla_k' \frac{d  \omega_{k-\frac{eEt}{\hbar}}}{dt} = \nabla_k' \nabla_k \omega_{k-\frac{eEt}{\hbar}}\frac{e E}{\hbar}
=  \hbar^{-2}  \nabla_k' \nabla_k \epsilon_{k-\frac{eEt}{\hbar}} e E = -\frac{1}{m^{\ast}} eE, 
\end{eqnarray}  where $m^{\ast} =  \hbar^{2}( \nabla_k' \nabla_k \epsilon_{k+\frac{eEt}{\hbar}})^{-1}$ is the effective mass matrix.

We have studied the acceleration of the wavepacket in presence of electric field. We may easily generalize it to a magnetic field $B \hat z$ in the $\hat z$ direction. We just note that

\begin{eqnarray}
v_g(t) &=& \nabla_k' \omega_{k(t)}.\\ 
\frac{d v_g(t)}{dt} &=& \nabla_k' \frac{d  \omega_{k(t)}}{dt} = \nabla_k' \nabla_k \omega_{k(t)}\dot{k}(t)
=  \hbar^{-1}  \nabla_k' \nabla_k \epsilon_{k(t)} \dot{k} = \frac{\hbar}{m^{\ast}}\dot{k}. 
\end{eqnarray} Then using same argument as in Eq. (\ref{eq:kcyclotron} and  \ref{eq:cyclotron}), we get,

\begin{equation}
\frac{d v_g}{dt} = - \frac{e}{m} v_g \times B.
\end{equation}  

Thus we derived wavepacket dynamics of an electron in periodic potential. Of course, the whole treatment in this paper is in the absence of a resistance. The wavepacket in reality scatters of phonons and impurites. The present paper only details the dynamics of electron wavepacket between collisions with lattice.

\end{document}